\documentclass[twocolumn,showpacs,preprintnumbers,amsmath,amssymb,floatfix,nofootinbib]{revtex4}

\usepackage{graphicx}
\usepackage{bm}

\begin{document}


\title{Excited-State Density Distributions of Neutron-Rich Unstable Nuclei}

\author{J.~Terasaki}
 \affiliation{School of Physics, Peking University, Beijing 100871, P.~R.~China}
\author{J.~Engel}
 \affiliation{Department of Physics and Astronomy, University of North Carolina,
Chapel Hill, NC 27599-3255}

\date{\today}

\begin{abstract}
We calculate densities of excited states in the
quasiparticle random-phase approximation (QRPA)
with Skyrme interactions and volume pairing.
We focus on low-energy peaks/bumps
in the strength functions of a range of Ca, Ni, and Sn isotopes for $J^\pi =
0^+, 1^-$, and $2^+$.  We define an ``emitted-neutron
number", which we then use to distinguish localized states from
scattering-like states.  The degree of delocalization either increases as the
neutron-drip line is approached or stays high between the stability line and the drip line.  In the  $2^+$ channel of Sn, however, the low-lying states, not even counting surface vibrations, are still fairly well
localized on average, even at the neutron drip line.
\end{abstract}

\maketitle


The structure of excited states in exotic nuclei has been much studied 
recently, both in nuclei between Li and O \cite{Zin97, Par03,
Nak99, Try03} and in heavier Ca \cite{Har04} and Sn \cite{Adr05}
isotopes.  Low-energy strength is often enhanced, and theorists have
tried to understand the mechanisms responsible.  Besides strength
functions, they have examined transition densities, which can be
measured and provide information on excited-state structure.   Our
earlier paper \cite{Ter06} contains an investigation of strength
functions and transition densities in a large range of medium-heavy
spherical nuclei, from one drip line to the other.

Transition densities tell us about where in the nucleus transitions occur; they
reflect the spatial distribution of products of single-particle and single-hole
wave functions and, as a consequence, are always localized, though they can be
very extended near the neutron drip line.  In nuclei near stability large
transition strength implies collectivity, which in turn tends to cause
localization (as we shall see in giant resonances).  Near the drip line,
however, the familiar connection between strength and collectivity is less
systematic at low energies.  Reference~\cite{Ter06} shows many examples of
behavior like that characterizing light halo nuclei and predicted for certain
$0^+$ states in heavy nuclei in Ref.\ \cite{Ham96}: large strength coming from
non-collective transitions to very spatially-extended single-particle states.
Those excited states are usually unbound, and may not even be quasibound, that
is they may be completely delocalized scattering states \cite{Ham96}. 
 
In this letter we suggest a measure of localization
and then investigate the extent to which the strong low-lying excited states
are localized, independent of their collectivity.  To uncover changes in
structure near the drip line, we track the degree of localization as $N$
increases in $J^{\pi} = 0^+,1^-$, and $2^+$ channels.  Localization, which we
analyze by examining \emph{diagonal} density distributions of the excited states,
has not been systematically studied before and offers a new window into
excitations in exotic neutron-rich nuclei.

We can adopt ideas from simple one-particle quantum mechanics to distinguish
localized states from extended ones.  If the energy of a single particle in,
e.g., a square-well potential is negative (Chap.~3 of Ref.~\cite{Mes59}) then
the tail of its wave function decays exponentially, and if the energy is
positive the tail oscillates.  In certain small positive-energy windows,
however, the amplitude of the oscillating tail is much smaller than that of
the wave function inside the potential.  Such states are sometimes called 
``quasibound'', and the solutions that exhibit no enhancement inside
are sometimes referred to as ``scattering states.''

Of course we are dealing with many-body quantum mechanics and the
asymptotic wave function depends on many coordinates.  The tail of
the one-body density, however, should still give us a
measure of localization. In this letter, we use the Skyrme-QRPA wave functions
(with the parameter set SkM$^{\ast}$ and volume pairing) from Ref.\ \cite{Ter06}, in
which we discussed transition densities, to obtain the
diagonal density of excited-state $k$:
\begin{eqnarray}
\rho_k^q(\bm{r}) &=& \langle k| \sum_i^{N {\rm or} Z}\delta(\bm{r}-\bm{r}_i)
|k\rangle \nonumber\\
 &=&
\sum_{KK^\prime L\,;\,q}\left(-\psi_K^\ast(\bm{r})\psi_{K^\prime}(\bm{r})u_K u_{K^\prime}
\right.\nonumber \\
&&\left.+\psi_{\bar{K}}^\ast(\bm{r})\psi_{\bar{K}^\prime}(\bm{r})v_K v_{K^\prime}\right)
\nonumber\\
&&\times\left(X_{LK^\prime}^k X_{KL}^k + Y_{KL}^k Y_{LK^\prime}^k\right)
+ \rho_0^q(\bm{r})~,
\label{eq:den_ex}
\end{eqnarray}
where $q$ is proton or neutron, $\bm{r}_i$ is the position operator of
particle $i$,
$\psi_K(\bm{r})$ is a single-particle wave function in the canonical
basis, and $u_K$ and $v_K$ are the associated occupation amplitudes.  The QRPA
amplitudes $X^k_{KK^\prime}$ and $Y^k_{KK^\prime}$ are assumed real,
$\bar{K}$ denotes the state conjugate to $K$, and we have used the convention
$\psi_{\bar{\bar{K}}}(\bm{r})=-\psi_K(\bm{r})$.
The summation is taken only for protons or neutrons,
and in our numerical calculations we will
use the Hartree-Fock (-Bogoliubov) approximation to the ground-state density
$\rho_0^q(\bm{r})$.
To obtain Eq.\
(\ref{eq:den_ex}) we have used the excited-state creation operator
\begin{equation}
O_k^\dagger = \frac{1}{2}\sum_{KK^\prime}\left(X_{KK^\prime}^k
a_K^\dagger a_{K^\prime}^\dagger
-Y_{KK^\prime}^k a_{K^\prime} a_K \right),
\label{eq:okdagger}
\end{equation}
\begin{equation}
X^k_{K^\prime K} = -X^k_{KK^\prime},
Y^k_{K^\prime K} = -Y^k_{KK^\prime},\ K<K^\prime, \label{eq:ext_X_and_Y}
\end{equation}
where $a^\dagger_K$ and $a_K$ are the creation and annihilation operators
for quasiparticles in the canonical state $K$.

If, as in all the calculations presented here, the ground state is spherically
symmetric, the number of particles between two radii is just the integral over
that well-defined region of $\rho_k^q(r)r^2$. We define the ``emitted-neutron
number'' associated with the state $k$ as \begin{equation}
N_e(k)=4\pi\int_{r\geq r_c} dr\;r^2(\rho_k^n(r)-\rho_0^n(r)), \end{equation}
where $r_{c}$ denotes the radius at which the density $\rho_k^n(r)$
begins to develop
a scattering tail.  If $N_e(k)$ is close to or larger than 1, then the excited
state is scattering-like (not localized), and if $N_e(k)$ is appreciably
smaller  than 1, then the excited state qualifies as quasibound (localized).
Intermediate cases are also possible, of course. If the pairing gap is not zero
$\rho_k^n(r)$ does not integrate to the correct particle number\footnote{The
QRPA does not guarantee the conservation of a quantity that is 
of second order or higher in the $X$'s and $Y$'s.}; we therefore multiply $\rho_k^n(r)$ by a constant (at
most a few percent from unity) to normalize it correctly.

Rather than focus on individual states, we want to examine peaks in the
strength function, which
can encompass several discrete states.  We therefore define a strength-weighted
average of the emitted-neutron number for the strongly excited states $k$ within 
a given peak:
\begin{equation}
\bar{N}_e = \frac{1}{2}\left( \bar{N}_e^{\text{IS}}+\bar{N}_e^{\text{IV}}
\right) ,
\end{equation}
\begin{equation}
\bar{N}_e^{\text{IS}} =
\frac{\sum_k S_k^{\text{IS}}N_e(k)}{\sum_k S_k^{\text{IS}}},
\bar{N}_e^{\text{IV}} =
\frac{\sum_k S_k^{\text{IV}}N_e(k)}{\sum_k S_k^{\text{IV}}},
\end{equation}
where $S_k^{\text{IS}}$ and $S_k^{\text{IV}}$ denote isoscalar and isovector
transition strengths to the state $k$.
The number of terms in the sums is between 1 and 10.
We average the isoscalar and isovector quantities  because both channels
appear in low-energy 
strength-function peaks, particularly near the drip line.  The giant resonances do not have this property, however, and for them
we will use $\bar{N}_e^{\text{IS}}$ or $\bar{N}_e^{\text{IV}}$ as measures.

\begin{figure}[tbh]
\includegraphics[width=0.45\textwidth]{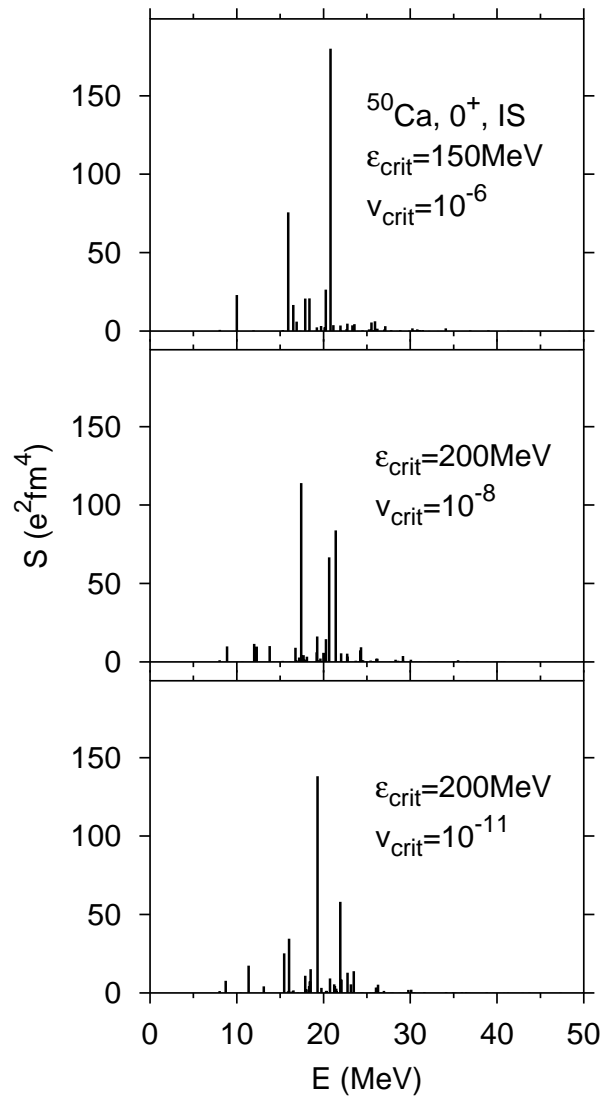} \caption{\label{fig2:str_is0+}
Strength distributions in the isoscalar 0$^+$ channel of $^{50}$Ca,
with a box size 20 fm.
The neutron chemical potential is $-$6.61 MeV.
}
\end{figure}

Before investigating the dependence of emitted-neutron number on $N$ and $Z$, we
need to see whether we can calculate it reliably.  The tail of the nuclear
density is extraordinarily sensitive, and converges slowly both with the size
of our spatial ``box" and the number of canonical states in our basis.  Figure
\ref{fig2:str_is0+} shows the isoscalar $0^+$ strength distribution for
$^{50}$Ca in a few versions of our calculation.  We have varied two cutoff
parameters: an upper limit $\varepsilon_{\rm crit}$ on canonical
single-particle energies (for protons, which have a vanishing paring gap), and
a lower limit $v_{\rm crit}$ on occupation amplitudes (for neutrons, which have
a non-vanishing gap).  We confirmed that when $\varepsilon_{\rm crit}=250$ MeV
and $v_{\rm crit}=10^{-14}$, the strength distribution is almost identical to
that with $\varepsilon_{\rm crit}=200$ MeV and $v_{\rm crit}=10^{-11}$, shown
in the bottom panel.  In Refs.\ \cite{Ter05,Ter06} we folded the strength to
account for the 20-fm box radius, and the values $\varepsilon_{\rm crit}=150$
MeV and $v_{\rm crit}=10^{-6}$ were sufficient for convergence of the strength
function and transition densities to the level of accuracy we required.  
Figure \ref{fig2:str_is0+} seems to suggests, however, that larger spaces are
required to reproduce the tiny part of the excited-state
density that lies significantly outside the nucleus.

\begin{figure}[t]
\includegraphics[width=0.45\textwidth]{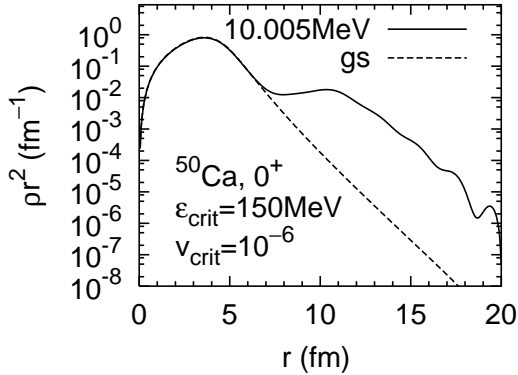}
\caption{\label{fig2:den_ex_is0+}
Neutron-density distribution of
the low-energy state with 
non-negligible strength
in $^{50}$Ca (see Fig.~\ref{fig2:str_is0+}), calculated with $\varepsilon_{\rm crit}=150$ MeV,
$v_{\rm crit}=10^{-6}$, and a box radius of 20 fm.
The excitation energy is at the top of the figure.
The ground-state density distribution (gs) is also drawn.
In this case, $r_c=7$ fm.
}
\end{figure}

\begin{figure}[t]
\includegraphics[width=0.45\textwidth]{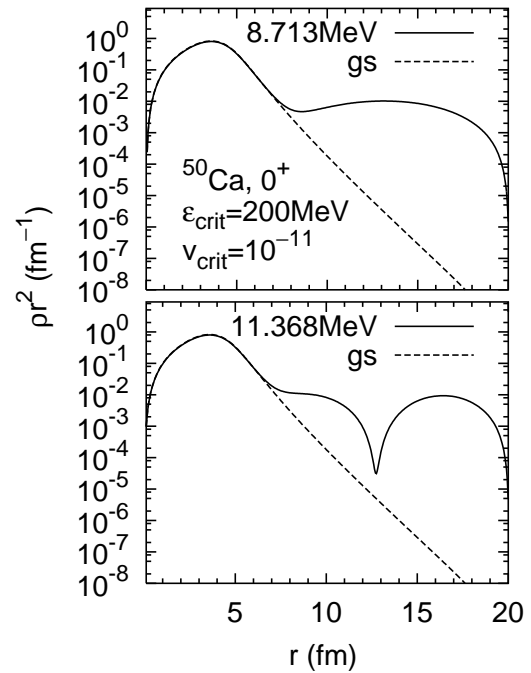}
\caption{\label{fig2:den_ex_e200_v-11}
The same as Fig.~\ref{fig2:den_ex_is0+} but for
$\varepsilon_{\rm crit} = 200$ MeV and $v_{\rm crit} = 10^{-11}$.
}
\end{figure}

To examine the issue more closely, we show in Fig.~\ref{fig2:den_ex_is0+} the
neutron density distribution associated with a low-lying state in the top  part
of Fig.\ \ref{fig2:str_is0+}; that distribution was obtained with the cutoff
parameters we used extensively in Ref.\ \cite{Ter06}.  In
Fig.~\ref{fig2:den_ex_e200_v-11} we show the neutron densities associated with
the corresponding states in the bottom part of Fig.\ \ref{fig2:str_is0+},
obtained in a larger space.  The tails of the more accurate densities are
noticeably different (and more realistic).  Yet improving the description this
way has a relatively small effect on the emitted-neutron number.  In the
smaller space the $\bar{N}_e$ of low-energy states is 0.84, with individual
$N_e(k)$ ranging from 0.77  to 0.90, and in the larger space it is 0.89, with
$N_e(k)$ ranging from 0.85 to 0.92.  These ranges, 10\% or less of the average,
are not only small, but also typical of cases with $\bar{N}_e$ near 1.
Changing the box radius to 25 fm fragments the strength by adding new states,
so that the $N_e(k)$ cannot really be tracked as the box size changes. The
average $\bar{N}_e$ over the states in a certain energy range, however, turns
out to be a fairly robust measure of localization; with a box size of 25 fm,
$\bar{N}_e=0.86$  for 5 sample states.  Perhaps the stability of $\bar{N}_e$,
an integrated quantity, is not so surprising given that the smaller space is
large enough to describe the wave functions extremely well for $r$ {\it less
than} $r_c$.  In any event, we can conclude here --- for the $0^+$ bump
below the giant resonance in $^{50}$Ca ---  that the states are
scattering-like. Nearly a full neutron is far outside the nucleus, no matter
how we vary parameters. 

To set a benchmark of sorts, we have looked at densities for giant resonances
in the isoscalar and isovector $0^+$, $1^-$, and $2^+$ channels of several Ca,
Ni, and Sn isotopes near stability.  $\bar{N}_e^{\text{IS}}$ and
$\bar{N}_e^{\text{IV}}$ are always on the order of 0.5 or less. The isoscalar
giant dipole resonances have the largest values.  The smallest belongs to the
isovector giant quadrupole resonance of $^{58}$Ni, with $\bar{N}_e^{\text{IS}}
= 0.01$.  The reasons for this very small value are that the main components of
the neutron excitation involve deep hole states and particle states near the
Fermi surface, and that protons contribute appreciably.

\begin{figure}[h]
\includegraphics[width=0.45\textwidth]{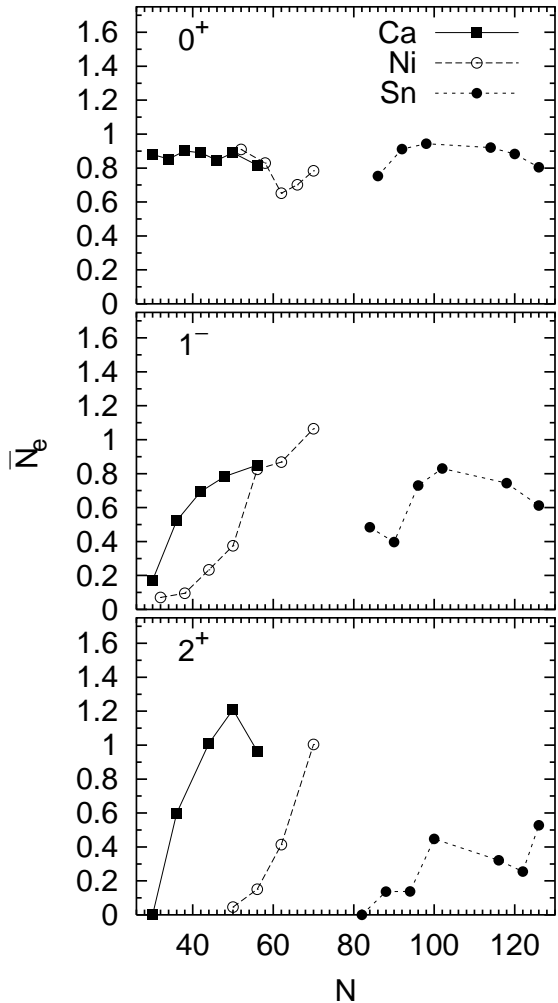}
\caption{\label{fig2:Ne}
$\bar{N}_e$ in low-lying strength-function peaks, versus
$N$, for the $0^+$, $1^-$, and $2^+$ channels of a sequence of Ca, Ni, and Sn
isotopes.
In the  $2^+$ channel, the lowest-energy states (surface vibrations), carrying
the largest strengths, are not included.
In Ni the $1^-$ curve has more points than those of the other $J^\pi$ channels because the low-energy
bump first appears at smaller $N$.}
\end{figure}

Having established a good measure of localization and calibrated it for typical
collective resonances, we are now in a position to examine the behavior of
low-lying excitations in neutron-rich nuclei.  In each channel we look at
low-energy bumps in a range of Ca, Ni, and Sn isotopes.  The resulting
$\bar{N}_e$, plotted versus $N$, appear in Fig.~\ref{fig2:Ne}.  For $0^+$
states they are more or less constant, at least in Ca and Sn, and since the
constant is greater than 0.8, the wave functions in the bumps are not
localized.  The $\bar{N}_e$  in the $1^-$ channel in Ca and Ni show a clear
increase with $N$, from 0.1 to about 1; those states are localized near
stability and become more scattering-like towards the drip line.  The $1^-$
states in Sn are  harder to interpret; the curve has a maximum of about 0.8
around $N=100$, but shows no clear trend in one direction or the other.  The
$2^+$ states in Ca and Ni mirror the behavior of the $1^-$ states.  In the
$2^+$ channel of Sn, $\bar{N}_e$ increases with $N$ but the low-lying states
never get less localized than typical giant resonances.  By our criterion,
those states are localized, on average, even at the drip line, though they
fluctuate within a given peak in a way that cannot be seen in the average (some
individual states have $N_e(k) > 0.8$).  We should note that the states used to
determine $\bar{N}_e$ do \emph{not} include low-lying surface quadrupole
vibrations, which usually have $N_e(k) < 0.1$.

In Ref.\ \cite{Ter06} we found that the strength to low-lying bumps grew wth
$N$ in essentially every channel and every isotopic chain.  Though we sometimes
see similar behavior in $N_e$, it is not universal.  The $0^+$ states, for
instance, are not localized even near stability.  And the $2^+$ states in Sn,
while they become more scattering-like with increasing $N$, do not always
delocalize beyond the level of giant resonances.  All this comes with a caveat,
however:  it is possible that many-particle many-hole states that admix with
QRPA excitations can alter $\bar{N}_e$ somewhat.  The one-particle one-hole
nature of our states means, for instance, that $\bar{N}_e$ is never much
greater than one.  Calculating the structure of many-particle many-hole states
is difficult, however, and it may be some time before a systematic study of the
kind reported here can be made.



To summarize quickly:  
giant resonances consistently have $\bar{N}_e$ around
0.5.  In low-energy peaks  $\bar{N}_e$ is more varied, and changes with neutron
number in ways that depend both on the isotopic chain and on multipolarity.
Stronlgy excited low-energy states near the drip line are sometimes scattering
states, but not always.

This work was funded in part by the U.S.\ Department of Energy under grant
DE-FG02-97ER41019.  We thank W.\ Nazarewicz for useful discussion and the
National Center for Computational Sciences at Oak Ridge National Laboratory and
Information Technology Services at University of North Carolina at Chapel Hill
for the use of their computers.  Parts of this research were done when one of
us (J.T.) was at the University of North Carolina and at RIKEN.

\end{document}